\documentclass{IEEEtran}

\usepackage{amsmath,amssymb,epsfig,psfrag,cite,subfigure}
\usepackage{graphicx}
\usepackage{epstopdf}

\usepackage{bm}

\usepackage{tcolorbox}

\usepackage{color}

\usepackage{pgfplots}
\usepackage{tikz}
\usetikzlibrary{shapes.geometric, arrows.meta, positioning}
\usepackage{varwidth}

\usepackage{hhline}
\usepackage{colortbl}
\usepackage{xcolor}
\usepackage{multirow}
\usepackage{booktabs}

\usepackage{array}
\usepackage{balance}
\newcolumntype{M}[1]{>{\centering\arraybackslash}m{#1}}  

\makeatletter
\@ifpackageloaded{xcolor}{}{
  \usepackage[table]{xcolor}
}
\makeatother

\usepackage{xr}

\makeatletter
\newcommand*{\addFileDependency}[1]{
  \typeout{(#1)}
  \@addtofilelist{#1}
  \IfFileExists{#1}{}{\typeout{No file #1.}}
}
\makeatother

\newcommand*{\myexternaldocument}[1]{%
    \externaldocument{#1}%
    \addFileDependency{#1.tex}%
    \addFileDependency{#1.aux}%
}

\myexternaldocument{supp}

\newcolumntype{V}{!{\vrule width 1.2pt}}  
\newcommand{\hthickline}{\noalign{\hrule height 1.2pt}}  

\newcommand{\rev}[1]{\textcolor{black}{#1}} 
\newcommand{\revv}[1]{\textcolor{black}{#1}} 

\author{Musa~Furkan~Keskin, Muralikrishnan  Srinivasan, Onur~G{\"u}nl{\"u},
Hui~Chen, Panagiotis Papadimitratos, Magnus Almgren,
Zhongxia Simon He, 
and~Henk Wymeersch\vspace{-2em}

\thanks{This work is supported by Chalmers Transport Area of Advance, the SNS JU project 6G-DISAC under the EU’s Horizon Europe research and innovation program under Grant Agreement No 101139130, the Swedish Research Council (VR) under Grants 2023-03821, 2023-05184 and 2024-04390, the KAW WAF program, the ZENITH Research and Leadership Career Fund under Grant 23.01, the Swedish Foundation for Strategic Research (SSF) under Grant ID24-0087, German Federal Ministry of Research, Technology and Space (BMFTR) 6GEM+ Transfer Hub under Grants 16KIS2412 and 16KISS005, ANRF/ECRG/2024/004438/ENS and 5G Use Case Lab.}
}

\begin{document}

\bstctlcite{IEEEexample:BSTcontrol}

\title{Multi-Domain Security for 6G ISAC: Challenges and Opportunities in Transportation}
\maketitle

\begin{abstract}
Integrated sensing and communication (ISAC) will be central to 6G-enabled transportation, providing both seamless connectivity and high-precision sensing. However, this tight integration exposes attack points not encountered in pure sensing and communication systems. In this article, we identify unique ISAC-induced security challenges and opportunities in three interrelated domains: cyber-physical (where manipulation of sensors and actuators can mislead perception and control), physical-layer (where over-the-air signals are vulnerable to spoofing and jamming) and protocol (where complex cryptographic protocols cannot detect lower-layer attacks). Building on these insights, we put forward a multi-domain security vision for 6G transportation and propose an integrated security framework that unifies protection across domains \rev{by leveraging existing ISAC measurements for lightweight cross-checks}.

\end{abstract}


\vspace{-5mm}
\section{Introduction}
The emergence of 6G wireless systems is redefining network design by fusing communication and sensing into a unified framework known as integrated sensing and communication (ISAC). ISAC enables simultaneous data exchange and environmental perception through shared spectrum and hardware.
\rev{Compared to independent sensing and communication subsystems,} such dual use of resources enhances spectral efficiency, latency and situational awareness, 
supporting autonomous driving and intelligent traffic management. \rev{In particular, ISAC alleviates spectrum congestion in dense vehicular networks with a unified waveform and facilitates low-latency exchange of massive onboard data via sensing-assisted communications \cite{ISAC_Veh_magazine_2025}.} However, with this innovation comes a shift in security considerations. 
Security in this context involves two interdependent aspects: \textit{communication security}, which protects the authenticity, integrity and confidentiality of transmitted data, and \textit{sensing security}, which safeguards the privacy of sensed entities and ensures the trustworthiness of sensing results  \cite{secureISAC_survey_2025}.

\begin{figure}
\centering
\centerline{\includegraphics[width=0.98\linewidth]{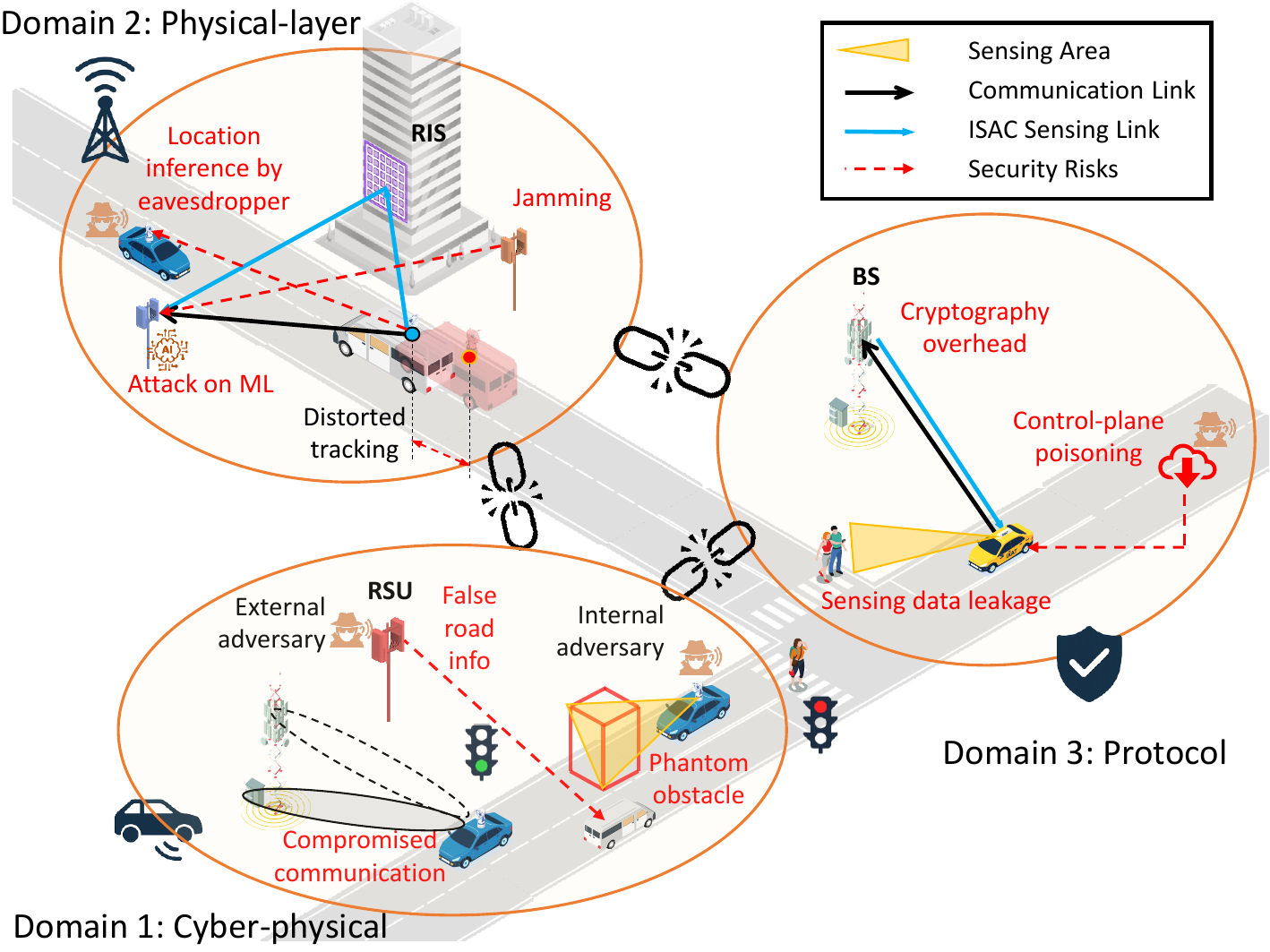}}
\caption{ISAC-induced vulnerabilities in transportation systems span three domains. The \textit{cyber-physical domain} forms the interface between physical environment and digital decision-making, where sensors (such as cameras and radar) observe the physical environment and provide input to autonomous decision systems, while actuators execute these decisions (e.g., steering). The \textit{physical-layer domain} governs the generation, propagation and reception of waveforms that simultaneously support communication and environmental sensing. The \textit{protocol domain} provides cryptographic protocols to ensure authenticity, integrity and confidentiality of user data and sensing information. \rev{The depicted links and attack vectors are intended to illustrate representative attack surfaces and cross-domain interdependencies.} A security breach in any domain can cascade into system-wide failures, calling for a coordinated multi-domain defense strategy.}
\vspace{-5mm}
\label{fig_vision}
\end{figure}

Prior research on secure  positioning has largely focused on authenticated peers and protected message exchanges while ISAC extends the challenge to \textit{uncooperative and passive targets}, shifting part of the attack surface into the environment itself. As a result, adversaries can exploit unencrypted over-the-air features, such as range, Doppler and angle, for unauthorized sensing and manipulate echoes to create ghost objects \cite{6G_ISAC_Security_2025,RIS_manipulation_magazine_2023}. In addition, due to joint use of  waveforms for sensing and communications, cross-layer coupling becomes tighter in ISAC, resulting in faster error propagation across layers (e.g., sensing spoofing modifies beam selection, which degrades link reliability). \rev{The waveform structure of ISAC with standardized reference signals and structured transmissions constitutes the root cause of many vulnerabilities, such as forced synchronization, advanced target mimicry, phantom target injection and radar overcrowding \cite{WLAN_sensing_Hasan_2025}.} These characteristics distinguish secure ISAC from classical secure positioning and demand a broader multi-domain defense tailored to unique threat surfaces introduced in ISAC, spanning \textit{(i) the cyber-physical domain} (sensor/actuator integrity and control-loop safety), \textit{(ii) the physical-layer domain} (robustness to spoofing/jamming/interference and environmental manipulation) and \textit{(iii) the protocol domain} (cryptographic measures to ensure authenticity, integrity and confidentiality for both data and sensing information).

 Existing works \rev{at} the intersection of ISAC and security either \textit{(i)} focus primarily on physical-layer security\rev{, privacy techniques and related trade-offs \cite{multifunc_6G_ISAC_2022,6G_ISAC_Security_2025}, or \textit{(ii)} investigate security functions in a layer-by-layer manner \cite{6G_ISAC_Security_2025,secureISAC_survey_2025} without devising a coordinated strategy that explicitly accounts for tight cross-layer coupling in ISAC. We argue that secure ISAC poses a \textit{cross-domain} problem, especially in safety-critical settings (e.g., attacks originating in sensing or communication subsystems can propagate across the perception-communication-control loop and lead to system-wide failures).} Our contributions are twofold: \textit{(i)} we systematically identify ISAC-specific security challenges and opportunities for transportation systems in the three domains and \revv{distinguish them from their counterparts} in standalone sensing and communication systems, and \textit{(ii)} \revv{bridging these domain-level insights,} we propose a \textit{multi-domain security approach} for ISAC in 6G that \revv{integrates} defenses across these domains \revv{into a unified security framework} \rev{through coupling-aware design (e.g., cross-domain authentication, key generation and anomaly detection)}. 


\begin{table*}[hbtp]
\centering
\scriptsize
\caption{Security Risks and Enhancements Across System Types and Individual Domains}
\renewcommand{\arraystretch}{1.7}
\setlength{\tabcolsep}{3pt}
\begin{tabular}{c V c | M{4.5cm} | M{4.5cm} | M{4.5cm}}
\hthickline
\textbf{} & \textbf{} & \textbf{Domain 1: Cyber-Physical} & \textbf{Domain 2: Physical-Layer} & \textbf{Domain 3: Protocol    } \\
\hthickline

\multirow{3}{*}{\rotatebox{90}{\textbf{Risks~~~~~}}} 
&  Pure Sensing 
&  Local tampering of onboard sensors  
&  External
jamming and spoofing of onboard sensors 
&  N/A \\
\hhline{~|*{4}{-}} 
&  Pure Communications 
&  Van Eck phreaking of sensor/actuator in-device communications
&  Pilot contamination, jamming, eavesdropping 
&  Leakage of meta data such as scheduling and network contexts\\
\hhline{~|*{4}{-}} 
& \cellcolor{red!5} ISAC 
& \cellcolor{red!5} Compromised ISAC infrastructure as additional sensor for actuator decisions
& \cellcolor{red!5} ISAC-amplified sensing privacy risks, compounding effects due to strong coupling in ISAC
& \cellcolor{red!5} Tight cross-layer coupling leads to inappropriate protocol-level decisions  \\
\hthickline

\multirow{3}{*}{\rotatebox{90}{\textbf{Enhancements~~~~~~} }} 
&  Pure Sensing 
&  Sensor redundancy (cross-sensing) 
&  Radar cross section (RCS) and micro-Doppler verification
&  
Consensus/attestation protocols using cross-verified sensing data
\\
\hhline{~|*{4}{-}} 
&  Pure Communications    
&  N/A
&  Secure beamforming, artificial noise (AN), artificial multipath, physical layer key generation 
&  
Using a secret key derived from physical parameters to initialize or refresh protocol-level security without a full cryptographic handshake
\\
\hhline{~|*{4}{-}} 
& \cellcolor{green!5} ISAC 
& \cellcolor{green!5} Infrastructure-based ISAC sensing for cross-verification of onboard sensors 
& \cellcolor{green!5} Geometry-coupled sensing feedback via ISAC, sensing-aided secure beamforming and AN injection
& \cellcolor{green!5} Context-aware cryptography, sensing-informed trust mechanisms \\
\hthickline
\end{tabular}
\label{tab_isac_security_table}
\end{table*}
\vspace{-5mm}
\section{Security Risks and Enhancements by ISAC}\label{sec_risks_enh}
In contrast to traditional systems, where sensing and communication operate independently, 
ISAC creates tightly coupled dependencies: a security compromise of one functionality can propagate into the other (see Fig.\,\ref{fig_vision}). This section systematically analyzes ISAC-induced risks and enhancements across the three domains, with a focus on ISAC-specific features that distinguish it from pure sensing and pure communication systems (\rev{see Table~\ref{tab_isac_security_table}, which shows how ISAC may introduce new security challenges and opportunities beyond standalone systems).} 
We categorize attacks according to the domain of the legitimate entity being targeted rather than the attacker's domain (i.e., the victim defines the domain). \revv{We consider both external and internal adversaries, with passive or active RF capabilities, targeting cyber-physical, physical-layer and protocol domains. Attackers may aim at privacy leakage, sensing/communication degradation, unsafe control actions and compromise of trust, authentication and key management.\footnote{\revv{We assume that standard cryptographic primitives and secure hardware used for key storage and attestation operate correctly. However, vehicles, roadside units (RSUs) and base stations (BSs) may still be compromised or malicious, which motivates the need for cross-domain verification rather than reliance on a single trusted layer, as will be explained in Sec.~\ref{sec_multi_level}.}}}



\vspace{-3mm}
\subsection{Domain 1: Cyber-Physical}

\subsubsection{Security Risks by ISAC}
In traditional systems, onboard sensors are self-contained and operate independently of the communication infrastructure. They can be targeted through local spoofing and tampering, but adversaries typically need to be physically close. 
Here, ISAC introduces two new attack surfaces. 
First, ISAC enables external sensing by infrastructure such as BSs and RSUs, whose measurements may feed directly into onboard perception and control algorithms. This effectively extends the sensor boundary beyond the vehicle, introducing two vulnerabilities: \textit{(i)} malicious RSUs/BSs (internal adversaries) and \textit{(ii)} attacks on their ISAC functionality by external adversaries through spoofing and jamming \cite{WLAN_sensing_Hasan_2025}. In both cases, corrupted infrastructure data can mislead onboard systems that lack cross-validation with local sensors, which may cause unsafe actuation (e.g., emergency braking triggered by phantom obstacles). Although such signal-level manipulation originates in the physical-layer, its impact manifests as a cyber-physical domain failure. Second, ISAC communication can rely on sensing results (e.g., for beam alignment) so that attacks against onboard sensors may compromise communication performance (e.g., reliability and latency) \cite{OnurSecureFeedbackedISAC}. \rev{Compromised sensing and communication can degrade trustworthiness in cooperative perception among vehicles, which may distort collective situational awareness and lead to unsafe control decisions \cite{CP_security_proc_IEEE_2025}.}

\subsubsection{Security Enhancements by ISAC}
ISAC also offers opportunities to improve resilience at the cyber-physical domain. First, 
by enabling independent, infrastructure-based sensing that allows for cross-verification of onboard sensor data. If, for instance, a vehicle’s radar is spoofed to hide a nearby obstacle, an ISAC-enabled BS might still detect it and notify the vehicle. This external input can help resolve conflicting observations. Infrastructure-based ISAC sensing also operates at elevated viewpoints over wide areas, providing a broad and complementary view of the environment. These perspectives make it possible to detect inconsistencies, such as a mismatch between a predicted trajectory and range-Doppler observations, and raise security flags. Second,  ISAC can  act as a fallback when onboard sensing fails. Infrastructure-based sensing may provide sufficient situational awareness to keep the system functioning safely. Moreover, inputs from ISAC-enabled RSUs/BSs can help vehicles validate claims by other vehicles (e.g., their geostamps), which can prevent attacks by malicious entities from impacting actuator decisions. Overall, the enhancement at the cyber-physical domain comes from a strengthened infrastructure with sensing in the sense that it can have a broad perception of the physical aspects of the system, enabling more resilient sensor/actuator loops.

\vspace{-3mm}
\subsection{Domain 2: Physical-Layer}\label{subsec:Domain2PL}

\subsubsection{Security Risks by ISAC}
At the physical layer, the first ISAC threat concerns sensing information leakage and privacy violations. As ISAC signals propagate through the environment, they reflect off objects like vehicles and pedestrians, and thus directly embed geometric information such as position and velocity (i.e., through range, angle and Doppler measurements). Therefore, unlike traditional communication systems where the outcome of sensing is transmitted via secured data (e.g., exchange of sensory data via V2X communications for cooperative perception), ISAC allows external adversarial receivers to extract environmental information directly from the received waveform (e.g., using unencrypted  pilots) without decrypting the payload. As an example, to undermine location privacy in spite of pseudonymity, unlinkability or even encryption that could be applied even on safety beacons and conceal location data, an eavesdropper can passively intercept  signals and estimate object trajectories \cite{WLAN_sensing_Hasan_2025}. 
%
%
In addition to the threats resulting from passive attacks, which fall under the umbrella of \textit{exploratory attacks} \cite{security_review_2021_arslan}, malicious entities can also launch active attacks against ISAC systems. Adversaries may inject false channel state information (CSI) via pilot contamination attacks, distort channel feedback and perform coordinated jamming, all of which degrade sensing accuracy and communication performance (e.g., CSI-based precoding) \cite{security_review_2021_arslan}. The structured nature of ISAC waveforms (e.g., standardized  pilots \rev{and synchronization procedures} in \rev{OFDM-based} Wi-Fi sensing \cite{WLAN_sensing_Hasan_2025}) makes them vulnerable to spoofing attacks, where adversaries inject false reflections to create phantom targets. These active signal manipulations can mislead both sensing (e.g., object detection) and communication (e.g., beam alignment, user tracking), leading to more severe consequences in ISAC than in pure sensing and communication. 
\rev{Besides conventional threats, reconfigurable intelligent surface (RIS)-assisted ISAC introduces  vulnerabilities when the assumed RIS configuration deviates from its actual RIS state due to adversarial reprogramming (e.g., by altering reflection angles to distort received signals \cite{RIS_manipulation_magazine_2023}), hardware impairments and geometry uncertainties. In parallel, AI-driven ISAC pipelines (e.g., learning-based beam prediction and cooperative sensing fusion) can be subject to adversarial attacks, including data poisoning during training and carefully crafted perturbations during inference \cite{dataPoisoning_2024}. Another threat can result from side-channel attacks that exploit potential leakage of sensing and localization information in ISAC systems due to power patterns and oscillator artifacts.}


\subsubsection{Security Enhancements by ISAC}
ISAC brings security enhancements at the physical layer in numerous aspects. First, coding-theoretic methods can enable secure communication rates to exceed the secrecy capacity of ISAC systems \revv{without feedback (i.e., the no-feedback baseline) by leveraging} sensing feedback \cite{OnurSecureFeedbackedISAC}. Similarly, joint coding and beamforming strategies can be developed for sensing-assisted secure communication. 
Relative to pure secure communications,  ISAC provides an implicit, geometry-coupled sensing feedback via echoes, which  
can be exploited to surpass no-feedback secrecy limits and to adapt codebooks/beamforming under mobility without explicit CSI overhead. 
%
Second, an ISAC BS can leverage backscattered downlink signals to sense both the user equipment (UE) and surrounding objects, and optimize its beamforming to focus energy on the UE while suppressing leakage to the directions of 
potential adversarial receivers \cite{secureISAC_TWC_2021}. 
ISAC also enables adaptive artificial noise (AN) injection in both spatial and temporal domains. Specifically, AN can be directed towards adversarial regions to degrade their signal-to-noise ratio (SNR) while preserving a required quality at the legitimate UE \cite{secureISAC_TWC_2021}. Sensing-aided AN injection not only strengthens \textit{data confidentiality} (by protecting user data from eavesdropping) but also improves \textit{sensing privacy} (by distorting pilots and thus impairing the ability of adversaries to perform sensing of UEs and objects).

\begin{figure*}
\centering
\centerline{\includegraphics[width=1\linewidth]{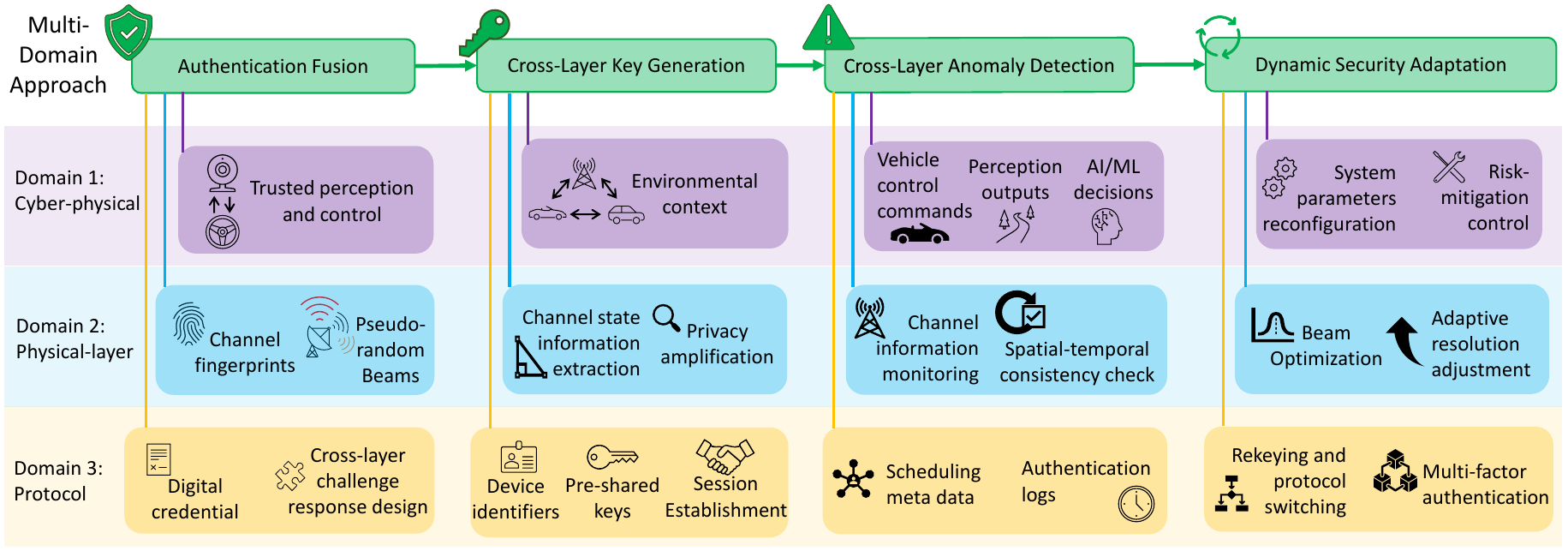}}
\caption{\rev{Multi-domain security approach for ISAC systems, illustrating how security functions can be coordinated across the cyber-physical, physical-layer and protocol domains. The four stages, namely, authentication fusion, cross-layer key generation, anomaly detection and dynamic security adaptation, form a logical security cycle. Each stage is connected to representative tasks in different domains and highlights the key aspects within the overall multi-domain defense paradigm.}} 
\label{fig_multi_domain}
\vspace{-0.2in}
\end{figure*}
\vspace{-3mm}
\subsection{Domain 3: Protocol}\label{sec_protocol_domain}

\subsubsection{Security Risks by ISAC}
As discussed in Sec.~\ref{subsec:Domain2PL}, in ISAC systems the attackers can sidestep encryption entirely by exploiting the analog sensing channel, extracting private information without violating payload confidentiality. Moreover, ISAC systems often rely on real-time feedback and cross-layer optimization, introducing tight coupling between layers. An adversary who manipulates lower-layer behavior (e.g., spoofed CSI, injected delay) can trigger protocol-level decisions such as handovers or route changes based on corrupted sensing inputs. 
A key challenge is to distinguish malicious manipulations from benign new nodes, for example, by checking consistency in transmit power, spatial alignment and timing. This kind of control-plane poisoning is amplified by the vertical integration of ISAC. In addition, post-quantum cryptography (PQC) 
and its resource demands pose new concerns for ISAC. The large key sizes and processing delays of PQC schemes may be incompatible with low-latency and real-time sensing requirements 
in the safety-critical transportation domain. 

\subsubsection{Security Enhancements by ISAC}
ISAC also opens up novel paths for strengthening higher-layer security through context-aware cryptography and sensing-informed trust mechanisms. ISAC sensing can support location-based authentication and context-aware key generation. For example, a shared key may be derived not only from the wireless channel but also from aligned sensing events (e.g., detecting the same target and confirming relative locations), providing an additional source of secrecy. 
As a further enhancement at higher layers, sensing information can improve trust management and anomaly detection. To create a hybrid trust model that combines digital authentication with physical evidence, devices whose trajectories or radio fingerprint  histories deviate from expected patterns can be flagged as suspicious even when their cryptographic credentials remain valid. ISAC further enables upper-layer systems to cross-check communication integrity using physical-layer cues, such as abrupt beam misalignment and inconsistent signal strength (which may indicate tampering/spoofing). In this way, cryptographic security is extended into the physical domain. 


\vspace{-3mm}

\section{Integrated Multi-Domain Secure ISAC}\label{sec_multi_level}
Sec.~\ref{sec_risks_enh} outlines defenses within individual domains, often designed to operate in isolation. As already mentioned, threats in ISAC-enabled transportation rarely stay confined to a single domain. For instance, control decisions in the cyber-physical domain can be detected at the physical layer \cite{huang2018combating}, while sensing data leakage can undermine cryptographic guarantees at higher layers. 
To address this interdependence, we introduce four cross-domain approaches that together form a security cycle (see Fig.\,\ref{fig_multi_domain}): \textit{(i)} authentication, \textit{(ii)} key generation, \textit{(iii)} anomaly detection and \textit{(iv)} security adaptation, ensuring coordination rather than replacement of domain-specific measures. \textit{Authentication fusion} combines physical-layer fingerprints with cryptographic methods, countering impersonation even when devices evade cryptographic encryption. The generation of \textit{cross-layer keys} transforms the randomness of the physical layer into a secret key, mitigating risks such as pilot contamination. \textit{Cross-layer anomaly detection} correlates sensing dynamics with protocol layer reports to determine inconsistencies caused by signal manipulation, AI poisoning, remote device spoofing, etc.  Finally, \textit{dynamic security adaptation} ensures that once anomalies are detected, the response mechanism is robust. 

\rev{
System-level deployment entails trade-offs in latency, overhead and implementation complexity. Multi-domain security should therefore operate selectively, reusing already-available ISAC observables for lightweight checks and invoking more costly protocol-level procedures such as strengthened authentication and rekeying only when inconsistencies are detected. Tighter sensing-communication coupling can accelerate cross-domain propagation of errors and the impact of attacks, but it also enables earlier cross-validation by correlating evidence across domains. The proposed mechanisms can be integrated incrementally into existing stacks without requiring a redesign of the underlying ISAC waveforms and protocols.
}

\vspace{-3mm}
\subsection{Authentication Fusion}
ISAC-enabled physical layer authentication (PLA) can leverage channel‐specific fingerprints, such as channel impulse response (CIR), angle and Doppler, that are difficult to forge over the millisecond time scales relevant to  transportation. In an ISAC transceiver, pilots already transmitted for joint radar/communication naturally expose these fingerprints, eliminating the need for additional signaling. The verifier builds the  features of the legitimate node during the offline acquisition/training  phase  and compares it with the ones freshly measured for every ISAC frame during the online authentication phase. 
To sharpen discrimination, ISAC hardware can actively shape the channel in real time by, e.g., randomly selecting a RIS pattern or switching to a pseudo-random beam. The goal is to ensure that only a prover physically present at the same location will exhibit channel variations consistent with the controlled perturbations of the  verifier \cite{TomasinPLA}. PLA is fast and context-aware, but by itself does not offer long-term key binding. Cross-layer fusion of PLA and protocol-domain cryptography can run both checks within a single ISAC frame: \textit{(i)} the transmitter appends a lightweight cryptographic key, generated using popular cryptographic algorithms, and \textit{(ii)} the verifier performs the PLA test on the induced feature. The packet is accepted only if both the digital token and the physical signature match. Cross-layer challenge-response designs of this form have been shown to significantly reduce the probabilities of impersonation and replay success, compared to the use of either layer alone, while adding negligible airtime overhead \cite{YongguCrossLayer}. In transportation scenarios such as intersection crossing and platooning, ISAC-enabled sensing features (e.g., Doppler, angle-of-arrival (AoA), or angle-of-departure (AoD)) provide richer fingerprints than conventional communication-only signals, making PLA-cryptography fusion especially effective. Through dual-factor handshakes, the run-time efficiency and  resilience of PLA can be combined with the established robustness of cryptographic methods to yield a scalable authentication framework for multi-domain secure ISAC, which can limit the impact of adversaries attempting to impersonate control/sensing nodes. 

\begin{figure}
    \centering
    \input{Figures2/PLA_R2}
    \vspace{-0.1in}
\caption{\revv{Authentication fusion case study: Expected complexity ratio $\rho(\gamma)$ for cross-layer authentication under different SNR regimes, subject to dual constraints (on false alarm and miss-detection probability). 
The ratio quantifies the average computational load relative to conventional HLA only. Results are shown as a function of the PLA threshold~$\gamma$, using a 16-element uniform linear array (ULA), 32-beam analog codebook over a $[-50^\circ,50^\circ]$ sector, and Alice located 
at~$10^\circ$. Eve's angle-of-arrival (AoA) is uniformly distributed over 
$[-60^\circ,60^\circ]$. The traffic mix consists of 50\% legitimate and 50\% 
illegitimate requests, with PLA cost set to 0.1\% of HLA cost. Shaded regions 
indicate the feasible set of~$\gamma$ satisfying both constraints  
(false alarm ($0.01$) and miss-detection ($0.1$) probability), and 
markers denote the optimal operating point~$\gamma^*$. At both SNR levels, the 
optimal~$\rho$ is approximately~0.5, corresponding to a~50\% complexity reduction from correctly rejecting illegitimate requests while accepting nearly all legitimate ones.}}
    \label{fig:PLAHLA}
    \vspace{-0.2in}
\end{figure}

\noindent \textit{Case study:} \revv{To demonstrate the concept of cross-layer fusion of PLA with higher-layer authentication (HLA), in Fig.\,\ref{fig:PLAHLA} we consider a two-stage system where PLA serves as a lightweight pre-check before conventional HLA. {At the carrier frequency of $28\,$GHz,} a BS estimates the AoA of each requesting user from uplink pilots and accepts it if the estimated AoA lies within a threshold~$\gamma$ of the registered value. {To ensure that computational savings do not come at the expense of authentication reliability, we impose dual constraints: a miss-detection bound  limiting the probability that an 
adversary passes PLA, and a false-alarm bound  ensuring that legitimate users are 
rarely rejected. These constraints define a feasible interval for~$\gamma$; within this interval we select the threshold minimizing expected complexity.} Only requests passing PLA undergo the computationally heavier HLA. {At the optimal operating point, the expected complexity is approximately~50\% of the HLA-only baseline at both high and low SNR, with the savings arising entirely from filtering out illegitimate requests. At very low SNR, the feasible region may vanish, correctly indicating that PLA gating requires sufficient estimation 
quality to be viable.}}


\vspace{-3mm}
\subsection{Cross-Layer Key Generation}
To turn intra-domain features into cross-domain keys suitable for session establishment and rekeying, we can bind physical/sensing context to cryptographic material by building on protocol domain enhancements in Sec.~\ref{sec_protocol_domain}. Beyond conventional bidirectional key exchanges, ISAC systems can derive secret keys directly from physical observables using location-dependent randomness as cryptographic material. While authentication fusion leverages these features to verify identity, cross-layer key generation transforms them into stable keys for ongoing secure communication. Since these features are tied to the environment and difficult to reproduce, they provide resilience against impersonation and replay attacks, similar to physical unclonable functions (PUFs) that exploit inherent manufacturing randomness. ISAC-based key generation operates through a layered pipeline that transforms physical randomness and cryptographic material into a unified key, as shown in Fig.\,\ref{fig_multi_domain}. At the \textit{physical layer}, devices such as vehicles or RSUs extract features from channel responses, AoAs/AoDs and multipath signatures, then quantize them into bit strings \cite{jiao2019physical}.
Reconciliation corrects mismatches across peers and privacy amplification compresses the aligned bits into uniformly random keys resilient to inference attacks. In parallel, the \textit{protocol layer} performs conventional cryptographic operations: authenticating devices, executing key agreement schemes and deriving session keys for secure communication. The two outputs are then fused through \textit{context binding} either via concatenation followed by hashing or by incorporating physical features such as CIR into the key derivation process. The result is a context bound key that integrates the unpredictability of the wireless environment with the structure of standard cryptographic protocols, remaining compatible with existing security infrastructures. 

\noindent \textit{Case study:} 
\rev{Fig.\,\ref{fig_KAP_vs_Alice_velocity_paper} plots the key-agreement probability (KAP) for the Alice-Bob and Alice-Eve links in a secure ISAC setup as a function of the velocity of Alice. In a challenging adversarial geometry where Eve is in close proximity to Bob, \revv{the legitimate and adversarial links exhibit a significant KAP gap, indicating strong spatial decorrelation that, after privacy amplification, translates into low residual leakage to Eve.}
At high velocities, Doppler-induced phase rotations decorrelate uplink and downlink channels in the Alice-Bob link, causing KAP to drop as velocity increases.}

\begin{figure}
    \centering
    \definecolor{mycolor1}{rgb}{0.85098,0.32549,0.09804}%
\definecolor{mycolor2}{rgb}{0.00000,0.44706,0.74118}%

\begin{tikzpicture}
[scale=1\columnwidth/10cm,font=\footnotesize]
\begin{axis}[
width=8cm,
height=4cm,
scale only axis,
xmin=0, xmax=30,
ymin=0, ymax=1,
xlabel={Velocity of Alice [m/s]},
ylabel={Key Agreement Probability},
axis background/.style={fill=white},
grid=both,
legend style={
  at={(0.02,0.68)},
  anchor=north west,
  legend cell align=left,
  draw=white!15!black,
  fill=white,
  fill opacity=0.8,
  text opacity=1
}
]

\addplot [color=mycolor1, dashdotted, line width=2.0pt, mark=o, mark options={solid, mycolor1}]
table[row sep=crcr]{%
0 0.996691871455577\\
2 0.995014955134596\\
4 0.975382932166302\\
6 0.938674033149171\\
8 0.905524861878453\\
10 0.838414634146341\\
12 0.823818293431553\\
14 0.76089588377724\\
16 0.721464019851117\\
18 0.669549104385423\\
20 0.656028368794326\\
22 0.622222222222222\\
24 0.613405146618791\\
26 0.561317876754118\\
28 0.562390158172232\\
30 0.535962877030162\\
};
\addlegendentry{Alice-Bob (Margin = $4$)}

\addplot [color=mycolor1, line width=2.0pt, mark=square, mark options={solid, mycolor1}]
table[row sep=crcr]{%
0 0.997164461247637\\
2 0.996510468594217\\
4 0.982494529540481\\
6 0.948618784530387\\
8 0.92707182320442\\
10 0.870121951219512\\
12 0.83609576427256\\
14 0.800847457627119\\
16 0.758064516129032\\
18 0.705991352686844\\
20 0.689716312056738\\
22 0.657057057057057\\
24 0.648713345302214\\
26 0.596095179987797\\
28 0.597539543057997\\
30 0.56554524361949\\
};
\addlegendentry{Alice-Bob (Margin = $8$)}

\addplot [color=mycolor2, dashdotted, line width=2.0pt, mark=o, mark options={solid, mycolor2}]
table[row sep=crcr]{%
0 0.0439508506616257\\
2 0.0483549351944167\\
4 0.0574398249452954\\
6 0.0430939226519337\\
8 0.0524861878453039\\
10 0.0518292682926829\\
12 0.0613873542050338\\
14 0.051452784503632\\
16 0.0694789081885856\\
18 0.0691785052501544\\
20 0.0555555555555556\\
22 0.0624624624624625\\
24 0.066427289048474\\
26 0.048810250152532\\
28 0.0527240773286467\\
30 0.0696055684454756\\
};
\addlegendentry{Alice-Eve (Margin = $4$)}

\addplot [color=mycolor2, line width=2.0pt, mark=square, mark options={solid, mycolor2}]
table[row sep=crcr]{%
0 0.0987712665406427\\
2 0.10568295114656\\
4 0.12417943107221\\
6 0.11767955801105\\
8 0.135359116022099\\
10 0.135365853658537\\
12 0.128299570288521\\
14 0.144673123486683\\
16 0.150124069478908\\
18 0.138974675725757\\
20 0.137115839243499\\
22 0.143543543543544\\
24 0.145421903052065\\
26 0.13849908480781\\
28 0.132396016403046\\
30 0.13631090487239\\
};
\addlegendentry{Alice-Eve (Margin = $8$)}

\end{axis}
\end{tikzpicture}
    \vspace{-0.1in}
    \caption{\rev{Cross-layer key generation case study: KAP vs. velocity of Alice in a secure ISAC scenario at $28 \, \rm{GHz}$ where  single-antenna Alice sends uplink signals to Bob and Eve (both equipped with $16$-element ULAs), and Bob responds in downlink to Alice. Two privacy amplification margins are considered under an $8$-bit cap, with the smaller one corresponding to more bits for key generation (hence, lower security risk but lower KAP for the legitimate Alice-Bob link). 
    \revv{The privacy amplification margin controls the trade-off 
between key rate and secrecy.}
    SNR is fixed at $40 \, \rm{dB}$ and Bob-Eve separation is $0.1 \, \rm{m}$. Both Alice-Bob and Alice-Eve channels have 8 paths. With increasing velocity of Alice, the Alice-Bob KAP drops due to Doppler-induced uplink/downlink channel decorrelation while the Alice-Eve KAP remains constant (evaluated conditional on Alice-Bob agreement).}}
    \label{fig_KAP_vs_Alice_velocity_paper}
    \vspace{-0.3in}
\end{figure}

\begin{table*}[t]
\centering
\caption{Mapping of Sec.~\ref{sec_risks_enh} Risks to Cross-Domain Mitigation}
\label{tab:risk_mitigation}
\begin{tabular}{p{3cm} p{4.5cm} p{7cm}}
\toprule
\textbf{Risk } & \textbf{Description} & \textbf{Mitigation } \\
\midrule
Cyber-physical: Remote device impersonation & Attackers mimic control/perception nodes to inject false data & \textbf{Authentication Fusion:} PLA-cryptography fusion validates both physical fingerprints and cryptographic authentication \\ 
\midrule
Cyber-physical: Communication degradation & Malicious nodes degrade sensing and communication quality & \textbf{Anomaly Detection:} cross-layer consistency checks expose spoofed and anomalous reports \\
\midrule
Physical: Privacy leakage & Reflections reveal user locations and trajectories  & \textbf{Dynamic Security Adaptation:} adaptive protocol/beam reconfiguration to limit leakage \\
\midrule
Physical: Pilot signal attack & Spoofed pilots corrupt channel estimation and beamforming & \textbf{Key Generation:} uses entropy from true environment, RSU cross-checks reject spoofed pilots \\
\midrule
Physical: RIS attack & Adversarial RIS manipulates reflections to mislead system & \textbf{Key Generation + Anomaly Detection:} mismatched entropy and context inconsistency exposes manipulation \\
\midrule
Physical: Reliance on AI & Poisoned models and adversarial samples misalign beams & \textbf{Anomaly Detection:} spatio-temporal smoothness and context correlation detect implausible AI-driven actions \\
\midrule
Protocol: Sensing not protected by cryptography & Over-the-air sensing bypasses cryptographic layers & \textbf{Authentication Fusion + Key Generation:} bind sensing features to cryptographic material for end-to-end security \\
\midrule
Protocol: Control-plane compromise & Attacks on rekeying, handovers, session management & \textbf{Dynamic Security Adaptation:} rekeying, protocol switching, fallback modes \\
\bottomrule
\end{tabular}
\vspace{-5mm}
\end{table*}

\vspace{-3mm}
\subsection{Cross-Layer Anomaly Detection}
In addition to secure key generation, ISAC also enables real-time cross-layer anomaly detection by correlating observables across the physical, protocol and cyber-physical domains, as shown in Fig.\,\ref{fig_multi_domain}. A \textit{unified detection engine} can expose threats along three complementary dimensions. First, \textit{cross-layer consistency checker} aligns physical features  with protocol metadata and control reports to reveal anomalies, for example, a Doppler-inferred velocity that contradicts a reported trajectory. \rev{In this context, the physical‑layer observables quantified in Fig.\,\ref{fig:PLAHLA} (such as the sensitivity of AoA‑based PLA pass/fail patterns to SNR) illustrate the types of temporal and geometric inconsistencies that a cross‑layer anomaly detection engine can exploit.} Second, a \textit{spatio-temporal analyzer} evaluates the natural evolution of multipath and Doppler features in temporal and spatial domains, flagging abrupt discontinuities and replayed patterns indicative of spoofing or RIS manipulation. 
Finally, \textit{context correlator} validates whether reflections and mobility states are consistent with environmental constraints such as road geometry and traffic flow. Here, mismatches across devices and with the map expose localized forgeries. Since most attacks perturb only one layer at a time, these complementary dimensions strengthen anomaly detection across layers and raise the bar for an adversary attempting to remain undetected.\revv{\footnote{\revv{In practice, lightweight checks of AoA/AoD consistency, multipath evolution and motion/trajectory/environment agreement can be performed locally at the vehicle or BS/RSU for time-critical operation or handled at the edge for broader correlation.}}}

\vspace{-3mm}
\subsection{Dynamic Security Adaptation}
Following anomaly detection, ISAC systems must also react quickly to contain threats before they propagate across domains. Dynamic security adaptation closes the loop by allowing the network to reconfigure itself in real time when inconsistencies appear. At the physical-layer, nodes may sharpen beampatterns, randomize RIS phases and improve radar resolution to reduce exposure toward suspicious directions. The protocol domain can respond by escalating authentication, adding multi-factor checks and switching to alternative observables for key generation when an existing observable becomes unreliable. At the cyber-physical domain, systems can raise trust thresholds and request confirmation from nearby sensors. These measures work best when coordinated. For instance, RSUs can gather anomaly reports and issue updated trust metrics and beam policies while infrastructure-less networks such as UAV swarms and V2V clusters adapt collaboratively through consensus and federated learning. Similarly, as shown in Fig.\,\ref{fig_KAP_vs_Alice_velocity_paper}, the mobility-induced key agreement fluctuations exemplify the run‑time stability checks that can trigger adaptive responses, such as switching key generation observables  and adjusting re‑keying frequency when legitimate reconciliation becomes unreliable.\revv{\footnote{\revv{Depending on the anomaly source and urgency, adaptation may remain local (e.g., reshaping beams, RIS randomization, rekeying) or be escalated to infrastructure-level measures such as protocol switching and fallback modes, while reusing existing ISAC measurements and protocol metadata to keep overhead manageable.}}} 
\vspace{-3mm}
\subsection{Summary and Levels of Integration} 
Table~\ref{tab:risk_mitigation} summarizes how the risks identified in Sec.~\ref{sec_risks_enh} map to the cross-domain defenses proposed in this section, highlighting the role of each approach within the overall security cycle.
%
\rev{While this paper mainly focused on full integration of communication and sensing (at the hardware and waveform levels), partial integration (e.g., site-level) is also an option. 
When risk tolerance is low and trustworthiness is uncertain, \textit{partial integration} can provide fault containment. For example, in ports and logistics hubs where jamming and spoofing are plausible and many third-party vehicles participate, dedicated roadside sensors (e.g., fixed radars and cameras) can maintain situational awareness even if BS/RSU communication is degraded. In such cases, V2X can be restricted to essential safety messages while auxiliary data (e.g., object and occupancy alerts with a reliability score) still supports site-level coordination. On the other hand, \textit{full integration} can be preferable when latency and performance requirements are stringent and cross-domain checks can run in real time. In highway platooning and scenarios with high-speed maneuvers, shared hardware and waveforms simplify alignment between sensing and V2X, which enables rapid reactions (e.g., emergency braking). Moreover, consistency checks can remain lightweight by reusing ISAC measurements and protocol metadata already produced in normal operation.}




\vspace{-0.1in}
\section{Open Research Challenges} 
We highlight three open security challenges for ISAC.
\vspace{-0.1in}

\subsection{Security of Distributed and Near-Field ISAC}
Distributed ISAC (DISAC) extends single-node ISAC by enabling multi-view sensing, wider coverage and improved robustness to blockage, but the required exchange of CSI, sensing results and synchronization signals also enlarges the attack surface and increases latency and privacy risks. Future work should focus on cooperative, privacy-preserving defenses, lightweight trust mechanisms and robustness against a small number of compromised nodes. Similar to DISAC, near-field ISAC with extremely large arrays introduces unique security risks due to spherical wavefronts and range-angle coupling. Adversaries may benefit from concentrated signal energy and perturb the focus point away from the intended target or UE. Possible mitigations include robust near-field beamforming under security and privacy constraints, accounting for focal point perturbations.  

\vspace{-0.1in}
\subsection{Role of AI in Multi-Domain ISAC Security}
AI is becoming a core part of ISAC through beam management, waveform design, cooperative perception and control, which means that model failures and attacks can now propagate across sensing, communication and decision-making. Adversaries may exploit this by poisoning training data, designing perturbations that lead to beam misalignment and learning device-specific signatures to improve spoofing and jamming. At the same time, AI can strengthen ISAC security through anomaly detection, distribution-shift monitoring and fast policy adaptation. Future work should therefore treat AI as an architectural security layer rather than a standalone tool, with emphasis on robust training, privacy-preserving learning and new threat models for complex AI-assisted ISAC stacks.

\vspace{-0.1in}
\subsection{Post-Quantum Cryptography and Non-Repudiation in ISAC}
PQC introduces \revv{important co-design} challenges for ISAC systems. The large key sizes, \revv{heavier} computation and extended handshake latencies of \revv{some post-quantum public-key mechanisms, such as key encapsulation and digital signatures, may create tension with the tight latency budgets of highly dynamic vehicular ISAC settings with frequent handovers and rekeying (unless mitigated through hardware acceleration and offloading), whereas symmetric-key operations are typically much less critical from a latency standpoint.}  These limitations call for new co-design principles where cryptographic exchanges are embedded within the ISAC frame structure rather than treated as separate network-layer processes. Key challenges include the allocation of compute and latency budgets across sensing, communication and security tasks, and the design of lightweight post-quantum primitives suited for short coherence times. Transportation systems also require non-repudiation for post-event accountability; here, permissioned distributed ledgers can asynchronously record hashes of ISAC \revv{events} without entering the real-time loop. PQC, physical-layer key generation and ledger-based auditing define an important security-latency tradeoff for future ISAC systems.
\vspace{-0.1in}


\balance 
\bibliography{Sub/references}
\bibliographystyle{IEEEtran}

\vspace{-0.15in}
\section*{Biographies}
\vspace{-0.05in}

\vskip -2.7\baselineskip plus -1fil

\begin{IEEEbiographynophoto}
{Musa Furkan Keskin} (furkan@chalmers.se) is a Senior Researcher with the Department of Electrical Engineering at Chalmers University of Technology, Sweden.
\end{IEEEbiographynophoto}

\vskip -2.7\baselineskip plus -1fil

\begin{IEEEbiographynophoto}
{Muralikrishnan  Srinivasan} (muralikrishnan.ece@iitbhu.ac.in) is an Assistant Professor with Indian Institute of Technology (BHU), Varanasi, India.
\end{IEEEbiographynophoto}

\vskip -2.7\baselineskip plus -1fil

\begin{IEEEbiographynophoto}
{Onur G{\"u}nl{\"u}} (onur.guenlue@tu-dortmund.de) is a Full Professor at TU Dortmund, Germany and a Guest Professor at Linköping University, Sweden.  
\end{IEEEbiographynophoto}

\vskip -2.7\baselineskip plus -1fil

\begin{IEEEbiographynophoto}
{Hui Chen} (hui.chen@chalmers.se) is a Research Specialist with the Department of Electrical Engineering at Chalmers University of Technology.
\end{IEEEbiographynophoto}

\vskip -2.7\baselineskip plus -1fil

\begin{IEEEbiographynophoto}
{Panagiotis 'Panos' Papadimitratos} (papadim@kth.se) is a Professor leading the Networked Systems Security (NSS) group at KTH Royal Institute of Technology, Sweden.
\end{IEEEbiographynophoto}

\vskip -2.7\baselineskip plus -1fil

\begin{IEEEbiographynophoto}
{Magnus Almgren} (magnus.almgren@chalmers.se) is an Associate Professor with the Department of Computer Science and Engineering at Chalmers University of Technology, Sweden.
\end{IEEEbiographynophoto}

\vskip -2.7\baselineskip plus -1fil

\begin{IEEEbiographynophoto}
{Zhongxia Simon He} (simon.he@chalmersindustriteknik.se) is a Researcher with Chalmers Industriteknik, Gothenburg, Sweden.
\end{IEEEbiographynophoto}

\vskip -2.7\baselineskip plus -1fil

\begin{IEEEbiographynophoto}
{Henk Wymeersch} (henkw@chalmers.se) is a Professor with the Department of Electrical Engineering at Chalmers University of Technology.
\end{IEEEbiographynophoto}

\end{document}